\documentclass[prl,twocolumn,showpacs,preprintnumbers,amsmath,amssymb,aps]{revtex4}

\usepackage{calc}
\usepackage{graphicx}
\usepackage{amsmath}
\usepackage{amssymb}
\usepackage{gensymb}
\usepackage{epsfig}
\usepackage{amsthm}
\usepackage{bm}
\usepackage{color}
\usepackage[toc,page]{appendix}
\usepackage{ragged2e}

\def\be{\begin{equation}}	
\def\ee{\end{equation}}
\def\arr{\begin{array}{rll}}
\def\ea{\end{array}}
\def\bea{\begin{eqnarray}}
\def\eea{\end{eqnarray}}

\begin{document}
\title{Impact of surface roughness on light absorption}
\author{V.Gareyan$^{1}$ and Zh.Gevorkian$^{1,2}$}
\affiliation{$^{1}$S.Matinyan Center for Theoretical Physics, Alikhanyan National Laboratory, Alikhanian Brothers St. 2,  Yerevan 0036 Armenia\\
$^{2}$Department of Theoretical Physics,Institute of Radiophysics and Electronics, Ashtarak-2 0203 Armenia}
\begin{abstract}
We study oblique incident light absorption in opaque media with rough surfaces. An analytical approach with modified boundary conditions taking into account the surface roughness in metallic or dielectric films has been discussed. Our approach reveals interference-linked terms that modify the absorption dependence on different characteristics. We have discussed the limits of our approach that hold valid from the visible to the microwave region. Polarization and angular dependences of roughness-induced absorption are revealed. The existence of an incident angle or a wavelength for which the absorptance of a rough surface becomes equal to that of a flat surface is predicted. Based on this phenomenon a method of determining roughness correlation length is suggested.
\end{abstract}

\maketitle

\section{ Introduction}

Surface roughness is an inherent property of all materials.
The scattering and reflection of electromagnetic waves from rough surfaces have been the subject of research by numerous papers(see Ref.\cite{simon2010} for a recent review). For a long time, it has been well known Refs.\cite{Brown85, McGurn87, McGurn96, Johnson99, Soubret01, Demir03, Navarette09} that,  the incident light is scattered diffusively if the characteristic size of roughness is of an order of the wavelength $\lambda$, and mostly specularly if it is much smaller.
 
Unlike reflection and scattering the roughness effect on absorption has received less attention.  However, in opaque systems, even from a scattering perspective in certain cases (for example, if one is interested in total scattered intensity), the calculation of absorptance is more convenient. Reflectance is directly found from the absorptance.
\color{black}
The effect of surface roughness on metal absorption at microwave frequencies was investigated in Refs.\cite{Morgan49}-\cite{Tsang06}. Early works on the absorption of electromagnetic waves in metals in optics focused on the absorption enhancement due to the excitation of plasmon polaritons on rough surfaces Refs.\cite{Fedders68, Ritchie68, Crowell70, Elson71, Kretschmann69, Juranek70,maradudin1,raether88,zayats05}. In the weak roughness case, the plasmon contribution to absorptance is negligible.

At optical frequencies roughness effect on metal absorption was studied
in Refs. \cite{Bergstrom08}-\cite{gevorkian22}. Below we extend the study reported in Ref.
\cite{gevorkian22}, focusing on the normal incidence onto the weakly rough metal surface, to the oblique incidence of different polarizations and dielectric films.
 It is shown that the oblique incidence case includes new important effects related to absorptance dependence on incident angle polarization, etc.
 The importance of the roughness effect on the absorption in nonmetals(Si for example)is associated with their wide use in microelectronics, chips, etc.,\cite{chips}. Particularly, the haze problem is very important in $Si$ wafers
\cite{haze}. In the case of semiconductor materials, both numerical and experimental studies have been reported to highlight the impact of surface roughness in light-matter interaction, specifically in the framework of solar cell efficiencies 
\cite{cell}.

The purpose of this paper is to develop a consistent perturbative approach to oblique incident light absorption by weakly rough surfaces characterized by Gaussian random profile function $h$ with RMS amplitude $\delta$  and correlation length $a$.  Note that other profile height distributions are possible as well. Fortunately, in real systems, the Gaussian distribution is often found as the proper one \cite{simon2010}.\color{black}

We want to develop an approach that describes wavelength regions from microwave to visible and infrared optics and is correct both for metals and dielectrics. 

One of the main problems of the perturbative approach in these systems is the choice of reference field \cite{carminati}.
 We address this issue by extending the boundary conditions for unperturbed fields to the actual surface profile and show that the accurate treatment of this problem significantly changes the first-order absorptance corrections \cite{gevorkian22}.

The paper is organized as follows. In Sec. 1, we set out our perturbation approach to the absorption by weakly rough surfaces. In Sec. 2, we calculate various contributions to the absorptance and derive the asymptotic expressions for the case of small-scale roughness. In Sec. 3, we discuss the results of our numerical calculations for silver and $Si$ films, and in Sec. 4 we summarize our results. The Appendix is devoted to the details of derivations of formulas.

\section{1. Initial Relations}
\label{sec-initial}

\begin{figure}[tb]
\vspace{0mm}
\centering
\includegraphics[width=0.8\columnwidth]{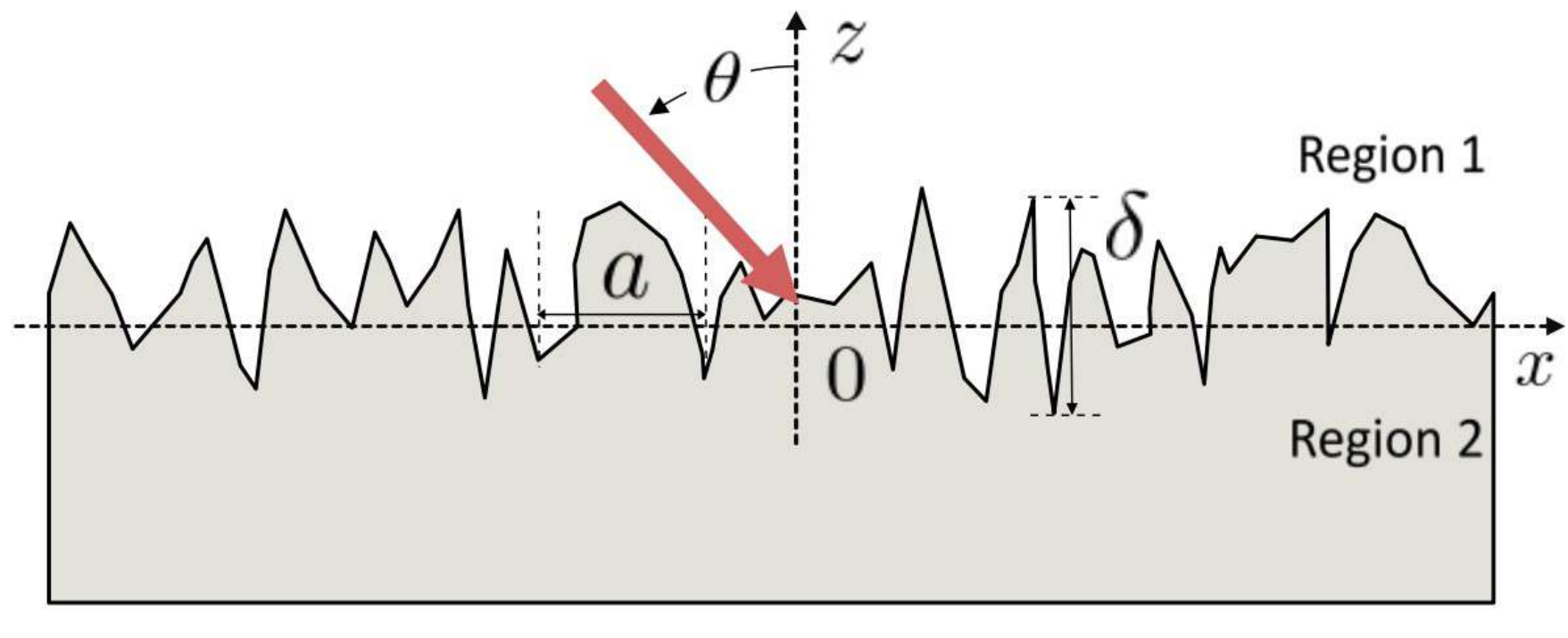}
\caption{Schematic illustration of the light incidence on the rough surface. Where, $\theta$ is the incidence angle of the light beam, $\delta$ is amplitude, and $a$ is the correlation length. Region 1 corresponds to the air and Region 2 to the material.  The effective dielectric constant of the medium consists of real and imaginary parts: $\varepsilon_{eff} = \varepsilon' + i \varepsilon''$.\color{black}}
\label{fig1}
\end{figure}

Consider the oblique incidence of a monochromatic wave $E,H\sim e^{-i\omega t}$  on the medium with rough surface, see Fig.\ref{fig1}

Maxwell equations have the form
\begin{equation}
{\bf \nabla\times E}=ik_0{\bf H},\quad {\bf \nabla\times H}=-ik_0 \varepsilon({\bf r}){\bf E}
\label{max1}
\end{equation}
where $k_0=\omega/c$,
\begin{equation}
\varepsilon({\bf r})=\Theta(z-h(x,y))+\varepsilon_{eff}\Theta(h(x,y)-z)
\label{roughsur}
\end{equation}
and $h(x,y)$ is the random profile of rough surface. Here we use Gaussian units.
The effective dielectric constant is introduced to describe both conducting and dielectric losses in metals and dielectrics from microwaves to visible and infrared optics \cite{krupka21}
\begin{equation}
\varepsilon_{eff}(\omega)=\varepsilon(\omega)+\frac{4\pi i\sigma(\omega)}{\omega}
\label{effdiel}
\end{equation}
Note that each of the two terms in Eq.(\ref{effdiel}) is important in different wavelength regions and for different materials. For example, for metals, the first term dominates in visible and infrared optics, while the second term is in microwaves.  Therefore one should substitute the proper expression for $\varepsilon$ in each wavelength region. \color{black} In most cases the roughness size is much smaller than the incident wavelength. Therefore one can expand Eq.(\ref{roughsur}) over $h$
\begin{align}
\varepsilon({\bf r},\omega)
&=\Theta[z-h(x,y)]+\varepsilon_{eff}(\omega)\Theta[h(x,y)-z]
\nonumber\\
&\approx\varepsilon_0(z,\omega)+\varepsilon_1({\bf r},\omega).
\label{dielcon}
\end{align}

where
\begin{align}
\varepsilon_0(z,\omega)=\left\{\begin{array}{cc}
         1, \mbox{~~~~for~} z>0 \\
         \varepsilon_{eff}(\omega), \mbox{for~} z<0
         \end{array}\right.
 \label{epszero}
 \end{align}
 is the permittivity for a smooth medium-air interface which, in the following, we refer to as the reference system,
\begin{equation}
\varepsilon_1({\bf r},\omega)=[\varepsilon_{eff}(\omega)-1]\delta(z)h(x,y),
\label{eps1}
\end{equation}
$\varepsilon_1$ is the perturbation due to small variations of $h$, $\delta(z)$ is the Dirac delta function. Excluding magnetic field from Eq.(\ref{max1}) one gets Maxwell equation for electric field
\begin{align}
&\bm{\nabla}^2{\bf E}-\bm{\nabla}(\bm{\nabla}\cdot{\bf E})+k_{0}^{2}\varepsilon({\bf r},\omega){\bf E}=0,
\label{max2}
\end{align}
It is convenient to decompose the electric field into parts ${\bf E}={\bf E}_0+{\bf E}_s$, namely reference and the scattered one due to roughness,  They correspondingly obey the following Maxwell equations
 \begin{align}
\nabla^2{\bf E}_0-\bm{\nabla}(\bm{\nabla}\cdot{\bf E}_0)+k_{0}^{2}\varepsilon_0(z){\bf E}_0=0,
\label{reffield}
\end{align}
and
\begin{align}
\nabla^2{\bf E}_s-\bm{\nabla}(\bm{\nabla}\cdot {\bf E}_s)+k_{0}^{2}\varepsilon({\bf r}){\bf E}_s=-k_{0}^{2}\varepsilon_1({\bf r}){\bf E}_0.
\label{scattfield}
\end{align}
The scattered fields can be obtained in a standard manner using the dyadic  Green's functions defined as \cite{maradudin1,maradudin2},
\begin{align}
\left[k_{0}^{2}\varepsilon_0(z)-\bm{\nabla}\bm{\nabla}+\nabla^2+k_{0}^{2}\varepsilon_1({\bf r})\right]\textbf{D}({\bf r},{\bf r}')=4\pi\delta({\bf r}-{\bf r}'),
\label{green}
\end{align}
where perturbation expansion over $\varepsilon_1$ is implied.

The reference field ${\bf E}_0$ can be chosen as the sum of incident and reflected plane waves in the air and a transmitted plane wave in the medium for a system with a smooth air-medium interface. However,  this is not a good choice for the $|\varepsilon'|\gg 1$ case because even a small change in the air-medium interface position leads to abrupt and significant field variations \cite{carminati}. To avoid this shortcoming, one  can modify the reference field by extending it up to the actual interface \cite{gevorkian22}:

\begin{align}
\tilde{E}_{0y}({\bf r})=\left\{\begin{array}{cc}e^{ik_xx-ik_{z}z}-re^{ik_xx+ik_{z}z},\quad \mbox{~for~} z>h(x,y),
\\
te^{ik_xx-ik_{z_{-}}z}, \quad \mbox{\qquad\qquad for~} z<h(x,y),
\end{array}\right.
\label{back}
\end{align}

where $k_x=k_0\sin\theta,k_z=k_0\cos\theta,k_{z_{-}}=k_{0}\sqrt{\varepsilon-\sin^2\theta}$, $\theta$ is the incidence angle and for brevity, we omit the index in $\varepsilon_{eff}$, $\varepsilon\equiv \varepsilon_{eff}$. Note that $\tilde{E}_{0x}=\tilde{E}_{0z}=0$ for our choice of s-polarization. Here, the incident wave amplitude $E_{\rm inc}$ is taken to be unity, while $r$ and $t$ are the standard Fresnel coefficients of reflection and transmission:
\begin{equation}
r=\frac{\sqrt{\varepsilon-\sin^2\theta}-\cos\theta}{\sqrt{\varepsilon-\sin^2\theta}+\cos\theta}, \quad
t=\frac{2\cos\theta}{\sqrt{\varepsilon-\sin^2\theta}+\cos\theta}.
\label{modt}
\end{equation}
Accordingly, the field decomposition now has the form ${\bf E}=\tilde{\bf E}_0+\tilde{\bf E}_s$, where the modified scattered field  is expressed through the dyadic Green's function $\textbf{D}({\bf r},{\bf r}')$ as
\begin{equation}
\tilde{\textbf{E}}_{s}({\bf r})=-\frac{k_{0}^{2}}{4\pi}\int d{\bf r}'\textbf{D}({\bf r},{\bf r}')\tilde{\textbf{E}}_{0}({\bf r}')\tilde{\varepsilon}_1({\bf r}').
\label{scatf}
\end{equation}
Here, $\tilde{\varepsilon}_1({\bf r})=(\varepsilon-1)\delta[z-h(x,y)]h(x,y)$ is the modified perturbation obtained from Eq.~(\ref{eps1}) by the replacement $z\rightarrow z-h(x,y)$ in the $\delta$-function in order to make it consistent with the extended boundary conditions Eq. (\ref{back}). Note that, while $\tilde{\varepsilon}_1({\bf r})$ and $\varepsilon_1({\bf r})$ coincide in the first order, the accurate choice of reference field leads to significant changes in roughness induced corrections as we show later in this paper.
 For the calculation of absorptance, one needs field values in the medium $z\leq h(x,y)$. Integration over $z'$ in Eq.(\ref{scatf}) is carried out through the $\delta$ function  in $\tilde{\varepsilon_1}$.\color{black}

For a monochromatic wave with frequency $\omega$, the power absorbed in a metal is given by  \cite{landau}
\begin{equation}
Q=\frac{\omega}{8\pi}\varepsilon''\int dV|{\bf E}|^2,
\label{loss}
\end{equation}
where integration is carried out over the medium volume and $\varepsilon''=Im\varepsilon+4\pi Re\sigma/\omega$ is the imaginary part of effective dielectric constant Eq.(\ref{effdiel}). The absorptance $A$ is obtained by normalizing $Q$ by the incident energy flux $Q_{\rm inc}=c|E_{\rm inc}|^{2}S_{0}cos\theta/8\pi$, where $S_{0}=L_{x}L_{y}$ is the normalization area. Using the above field decomposition, the absorptance averaged over the roughness configurations takes the form
\begin{equation}
A= \left \langle \frac{\varepsilon''k_{0}}{S_{0}\cos\theta}\int\! dV\! \left[|\tilde{{\bf E}}_0|^2+2\text{Re}(\tilde{{\bf E}}_0^*\cdot\tilde{\bf E}_s)+|\tilde{\bf E}_s|^2\right]\right \rangle.
\label{bulk2}
\end{equation}
 We evaluate all contributions to Eq.~(\ref{bulk2}) perturbatively, i.e., up to the order $\delta^{2}$. Specifically, we assume that the dimensionless parameters  $\delta/\lambda$, $\delta/d$ are small, but no restriction is imposed on the parameters $\delta/a,a/d$. Here $d$ is the penetration depth into the opaque medium (see below).\color{black}
\section{2. Electromagnetic Energy Loss in the Medium}
\label{sec-abs}

\subsection{ Reference field contribution}
\label{sec-ref}

Let us consider the first term in Eq.~(\ref{bulk2}) describing the reference field contribution:
\begin{equation}
A_{r}=\frac{\varepsilon''k_{0}}{S_{0}\cos\theta}\left\langle\int dV |\tilde{{\bf E}}_0|^2\right\rangle.
\label{bulbac}
\end{equation}
The integration over the medium volume can be presented as $\int dV=\int dx dy\int_{-\infty}^{h(x,y)}dz$, with $z=h(x,y)$ profile.  Note that in the determination of volume integral in \cite{gevorkian22} there is an unnecessary multiplier $\sqrt{1+h'^2}$\color{black}. Taking into account the extended boundary conditions  Eq.~(\ref{back}) , we have
\begin{align}
A_{sr}=\frac{\varepsilon''k_{0}|t|^{2}}{S_{0}\cos\theta}
\left\langle\int dxdy\int_{-\infty}^hdze^{-2\kappa_\theta k_{0}z}\right\rangle,
\label{bulbac2}
\end{align}
where we  adopted the standard notation $\sqrt{\varepsilon}=n+i\kappa$  for the complex refraction index as well as $\kappa_\theta=Im\sqrt{\varepsilon-sin^2\theta}$. Integrating over $z$ and expanding the integrand over  $h$ ,one gets
\begin{align}
A_{sr}=\frac{A_{s0}}{S_{0}}
\!\int\! dxdy\left[1+\frac{2\langle h^2(x,y)\rangle}{d^{2}}\right],
\label{bulbac3}
\end{align}
where $A_{s0}=\varepsilon''|t|^{2}/2\kappa_\theta cos\theta$ is the absorptance for an s-polarized wave by a smooth surface and $d=(k_{0}\kappa_\theta)^{-1}$ is penetration depth in the medium. Averaging over roughness configurations as $\langle h^2(x,y)\rangle=\delta^{2}$ , we finally obtain
\begin{equation}
A_{sr}=A_{s0}\left (1+\frac{2\delta^2}{d^2}\right ).
\label{bulkbackfin}
\end{equation}
\subsection{Interference term contribution}
\label{sec-int}

Next, consider the interference term
\begin{equation}
A_i=\frac{\varepsilon''k_{0}}{S_{0}\cos\theta}2Re\left\langle\int dV\tilde{{\bf E}_0^*}\tilde{{\bf E}}_s\right\rangle
\label{intterm}
\end{equation}

Up to the order $h^{2}$, the scattered field can be presented as a sum, $\tilde{\bf E}_s=\tilde{\bf E}_s^{(1)}+\tilde{\bf E}_s^{(2)}$, corresponding, respectively, to the lowest and first-order perturbation expansion of the Green function $\textbf{D}$ in Eq.~(\ref{scatf}). Accordingly, this contribution to the absorptance can also be split as $A_{i}=A_{i1}+A_{i2}$.
We start with the first contribution obtained by inserting the unperturbed Green's function $\textbf{D}_{0}$ corresponding to $\varepsilon_{1}=0$ in Eq.~(\ref{green}) into Eq.~(\ref{scatf}). One could think that since $\tilde{\bf E}_s^{(1)}\sim h$, the corresponding absorptance $A_{i1}$ would vanish after performing averaging over the roughness configurations. However, as we show below, the extended boundary conditions Eq.~(\ref{back}) for the modified reference field  $\tilde{{\bf E}}_0$ lead to a \textit{negative} contribution to the absorptance which balances out the excessive absorption increase due to scattered field penetration into the medium.

It is convenient to introduce two-dimensional Fourier transforms
\begin{equation}
D_{\mu\nu}(\vec{\rho}-\vec{\rho^{\prime}},z,z^{\prime})=\int\frac{d {\bf q}}{(2\pi)^2}d_{\mu\nu}({\bf q},z,z^{\prime})e^{i{\bf q}(\vec{\rho}-\vec{\rho^{\prime}})}
\label{fourier}
\end{equation}
Substituting expressions for ${\bf E_0,E_s}$ from Eqs.(\ref{back}),(\ref{scatf}), going to Fourier transforms over the plane coordinates $x,y$, this contribution can be represented in the form
\begin{eqnarray}
A_i^1=-\frac{k_0^3\varepsilon'' (\varepsilon-1)}{2\pi S_0 \cos\theta}|t|^2
Re\left\langle\int d{\vec \rho}d{\vec \rho'}\int_{-\infty}^hdz\right.\nonumber\\
\left.\int\frac{d{\vec q}}{(2\pi)^2}e^{i({\vec q}-k_x\vec e_x)({\vec \rho}-{\vec \rho'})}d_{yy}(\vec q,z,h(\vec{\rho^{\prime}}))\right.\nonumber\\
\left.e^{ik_{z_{-}}h(\vec{\rho^{\prime}})-ik_{z_{-}}^*z}h(\vec{\rho^{\prime}})\right\rangle \nonumber\\
\label{intterm2}
\end{eqnarray}
where $z$-coordinate is always in the medium $z<h(\vec \rho)$ and we used that $d_{yy}({\bf q},z,h_{+})=d_{yy}({\bf q},z,h_{-})$.
We note, that $h$ is included not only in the scattered fields but also in the material boundary. \color{black}
As was noted above bare Green's functions for the boundary problem are found in \cite{maradudin1}
\begin{equation}
d_{\mu\nu}({\bf q},z,z^{\prime})=\sum_{\alpha\beta}S_{\mu\alpha}^{-1}g_{\alpha\beta}(q,z,z^{\prime})S_{\beta\nu}
\label{deltage}
\end{equation}
where $g_{\mu\nu}(q, z,z^{\prime})$ is tabulated in \cite{maradudin1},\cite{maradudin2} and $3\times 3$ matrix is determined by the following elements $S_{xx}=S_{yy}=\hat{q_x}, S_{zz}=1, S_{xy}=-S_{yx}=\hat{q_y}, S_{xz}=S_{zx}=S_{yz}=S_{zy}=0$,
and $\hat{q_{x,y}}=q_{x,y}/q$. Using expressions for $S,S^{-1}$ and $g_{\mu\nu}({\bf q},z,z')$ one can make sure that
\begin{equation}
d_{yy}({\bf q},z,z')=\frac{g_{xx}(q,z,z')q_y^2}{q^2}+\frac{g_{yy}(q,z,z')q_x^2}{q^2}
\label{dyy}
\end{equation}
where
\begin{flalign}
g_{xx}(q,z,z')=-\frac{2\pi i k_1c^2}{\varepsilon\omega^2} \left[\frac{k_1+\varepsilon k}{k_1-\varepsilon k}e^{ik_1(z+z')}-e^{-ik_1|z-z'|}\right] \nonumber \\
g_{yy}(q,z,z')=-\frac{2\pi i}{k_1}\left[\frac{k_1+k}{k_1-k}e^{ik_1(z+z')}+e^{-ik_1|z-z'|}\right] \nonumber \\
\label{gxxyy}
\end{flalign}
and
\begin{eqnarray}
k(q)=\left\{\begin{array}{cc}
(\frac{\omega^2}{c^2}-q^2)^{1/2},q<\omega/c \nonumber \\
i(q^2-\frac{\omega^2}{c^2})^{1/2},\quad q>\omega/c \end{array}\right.\nonumber\\
k_1(q)=-\left(\varepsilon\frac{\omega^2}{c^2}-q^2\right)^{1/2}
\label{kandk}
\end{eqnarray}
Here we assume that the points $z,z'$ are in the medium. Substituting Eqs.(\ref{dyy}),(\ref{gxxyy}),(\ref{kandk}) into Eq.(\ref{intterm2}), expanding all the expressions on $h(\vec \rho)$, $h(\vec \rho')$ averaging by using $<h(\vec\rho)h(\vec\rho')>=\delta^2exp\left[-(\vec \rho-\vec\rho')^2/2a^2\right]$ and integrating alternately,one has
\begin{eqnarray}
A_i^1=\frac{4A_0\delta^2}{ad}Re[(\varepsilon-1)I^{2D}(\beta)]- \nonumber\\ -2A_0\delta^2k_0^2Re\left[(\varepsilon-1)\frac{\cos\theta-\sqrt{\varepsilon-\sin^2\theta}}{\cos\theta+\sqrt{\varepsilon-\sin^2\theta}}\right]
\label{intercontr1}
\end{eqnarray}
,where $\beta=k_0 a$ and the integral
\begin{eqnarray}
I^{2D}(\beta)=iexp(-\beta^2\sin^2\theta / 2) \int_{0}^{\infty} dr r  exp(-r^2/2) \times \nonumber\\ \times\left[ \frac{\sqrt{(\varepsilon \beta^2-r^2)(\beta^2-r^2)}}{\sqrt{\varepsilon \beta^2-r^2}+\varepsilon \sqrt{\beta^2-r^2}}f_{+}(r \beta \sin\theta) +\right.\nonumber\\ \left.
+ \frac{\beta^2}{\sqrt{\varepsilon \beta^2-r^2}+\sqrt{\beta^2-r^2}} 
f_{-}(r \beta \sin\theta) \right]\nonumber\\
\label{I2D}
\end{eqnarray}
with 
\begin{equation}
    f_{\pm}(x)=\frac{I_0(x) \mp I_2(x)}{2}
    \label{f}
\end{equation}
($I_n$ is the n-th modified Bessel function of the first kind).
Making use of $|\varepsilon|\beta^2\ll 1$ we get
\begin{eqnarray}
A_i^1=-\frac{A_0\delta^2\sqrt{2\pi}}{ad}Re\left[\frac{\varepsilon-1}{\varepsilon+1}\right]-\nonumber\\ -2A_0\delta^2k_0^2Re\left[(\varepsilon-1)\frac{\cos\theta-\sqrt{\varepsilon-\sin^2\theta}}{\cos\theta+\sqrt{\varepsilon-\sin^2\theta}}\right]
\label{intercontr}
\end{eqnarray}

 Now consider the second interference contribution $A_{i2}$. We expand the Green's function $\textbf{D}$, defined by Eq.~(\ref{green}),  to the first order in $h$, and  present the second-order scattered field in Eq.~(\ref{scatf}) as
\begin{align}
\tilde{\bf E}_s^{(2)}=\left (\frac{k_{0}^2}{4\pi}\right )^{2}\int & d{\bf r}'d{\bf r}''
\textbf{D}_{0}({\bf r},{\bf r}')\varepsilon_1({\bf r}')
\nonumber\\
&\times \textbf{D}_{0}({\bf r}',{\bf r}'')
\varepsilon_1({\bf r}'')\textbf{E}_{0}({\bf r}''),
\label{seconord}
\end{align}

Substituting Eq.(\ref{seconord}) into Eq.(\ref{intterm}) and evaluating $A_{i}^2$ in a similar manner, we obtain
\begin{equation}
A_{i}^2=A_0\frac{\sqrt{2\pi}\delta^2}{2ad}Re\left[\frac{(\varepsilon-1)^2}{i\kappa_\theta(\varepsilon+1)(\sqrt{\varepsilon-\sin^2\theta}+\cos\theta)}\right]
\label{intcontr2}
\end{equation}

Note that the formulae Eqs.(\ref{intercontr}),(\ref{intcontr2}) are correct provided that $|\varepsilon|\beta^2\ll 1$.

\subsection{Scattering term contribution}
\label{sec-scatt}

Consider now the scattered field contribution to the absorptance which is presented in the form
\begin{equation}
A_{sc}
= \frac{\varepsilon''k_{0}}{S_{0}} \left \langle \int\! dV \left (|\tilde{E}_{sx}|^2+|\tilde{E}_{sy}|^2+|\tilde{E}_{sz}|^2\right ) \right \rangle.
\label{bulkscat}
\end{equation}
After evaluating each term in the way outlined in the previous section, the result in the limit $|\varepsilon|\beta^2\ll 1$ can be presented as
\begin{equation}
A_{sc}=A_{0}\frac{\sqrt{2\pi}\delta^2}{2ad}\frac{|\varepsilon-1|^2}{|\varepsilon+1|^2}
\label{sctotal}
\end{equation}

Finally, the full absorptance for s-polarized wave  at oblique incidence is obtained by summing up all contributions: $A_s=A_{r}+A_{i}+A_{sc}$
 \begin{eqnarray}
 \frac{A_s}{A_{s0}}=1+\frac{2\delta^2}{d^2}-\frac{\delta^2\sqrt{2\pi}}{ad}Re\left[\frac{\varepsilon-1}{\varepsilon+1}\right] -\nonumber\\-2\delta^2k_0^2Re\left[(\varepsilon-1)\frac{cos\theta-\sqrt{\varepsilon-sin^2\theta}}{cos\theta+\sqrt{\varepsilon-sin^2\theta}}\right]+\nonumber\\
 +\frac{\sqrt{2\pi}\delta^2}{2ad}Re\left[\frac{(\varepsilon-1)^2}{i\kappa_\theta(\varepsilon+1)(\sqrt{\varepsilon-sin^2\theta}+cos\theta)}\right]+\nonumber\\+
 \frac{\sqrt{2\pi}\delta^2}{2ad}\frac{|\varepsilon-1|^2}{|\varepsilon+1|^2}
 \label{sfinal}
 \end{eqnarray}

 \subsection{p-polarization}

 In this case, reflection and transmission coefficients from smooth surfaces are determined as
 \begin{eqnarray}
 r_p=\frac{\varepsilon \cos\theta- \sqrt{\varepsilon-\sin^2\theta}}{\varepsilon \cos\theta+\sqrt{\varepsilon-\sin^2\theta}}, \nonumber\\
 t_p=\frac{2\sqrt{\varepsilon}\cos\theta}{\varepsilon \cos\theta+\sqrt{\varepsilon-\sin^2\theta}}
 \label{ppolar}
 \end{eqnarray}
The absorptance for the p-polarized wave is found analogously to s-polarization, see Appendix B.
 \begin{flalign}
 \frac{A_p}{A_{p0}}=1+\frac{2\delta^2}{d^2}-\frac{\delta^2\sqrt{2\pi}}{ad}Re\left[\frac{\varepsilon-1}{\varepsilon+1}\left( |n_x|^2-2|n_z|^2/\varepsilon \right)\right] \nonumber\\ -2\delta^2k_0^2 Re\left[\left(1-\frac{1}{\varepsilon}\right)\frac{\sqrt{\varepsilon-\sin^2\theta}-\varepsilon \cos\theta}{\sqrt{\varepsilon-\sin^2\theta}+\varepsilon \cos\theta}((\varepsilon-\sin^2\theta)|n_x|^2\right.\nonumber\\ \left.+2i\sin\theta\sqrt{\varepsilon-\sin^2\theta}Im(n_z^{*}n_x)-\sin^2\theta|n_z|^2)\right] \nonumber\\
 +\frac{\sqrt{2\pi}\delta^2}{2ad}Re\left[\frac{(\varepsilon-1)^2\cos\theta \sqrt{\varepsilon-\sin^2\theta}}{i\kappa_\theta(\varepsilon+1)(\sqrt{\varepsilon-\sin^2\theta}+\varepsilon\cos\theta)}  \times ( |n_x|^2 \right.\nonumber\\ \left. +\frac{2\tan\theta}{\varepsilon^2}n_x^* n_z  -\frac{\sin\theta}{\sqrt{\varepsilon-\sin^2\theta}} n_z^* n_x - \frac{2\sin\theta\tan\theta}{\varepsilon^2\sqrt{\varepsilon-\sin^2\theta}}|n_z|^2) \right] \nonumber\\+ 
 \frac{\sqrt{2\pi}\delta^2}{2ad}\frac{|\varepsilon-1|^2}{|\varepsilon+1|^2}\left[ 
 |n_x|^2+2|n_z|^2/|\varepsilon|^2 \right]\nonumber\\
 \label{pfinal}
 \end{flalign}
where $A_{p0}=\varepsilon''|t_p|^2\tau^2/2\kappa_{\theta}\cos\theta$ is the absorptance of p-polarized wave from smooth surface and $n_x=\tau^{-1}\sqrt{1-\sin^2\theta/\varepsilon}$, $n_z=\sin\theta/(\tau\sqrt{\varepsilon})$ together with $\tau=\sqrt{|1-\sin^2\theta/\varepsilon|^2+\sin^2\theta/|\varepsilon|}$.

%
%
Concluding this section, we note that the accurate choice of reference field Eq.(\ref{back}) ensures the small magnitude of first-order correction to the absorptance in the weak-roughness case.
Specifically, had we chosen the standard, rather than extended, boundary conditions for reference fields, the interference term giving negative contribution would be absent, and total absorptance  Eqs. (\ref{sfinal}),(\ref{pfinal}) would have increased, signaling a poor choice of basis set for the perturbation expansion.



\section{3. results and discussion}
\label{sec-num}
\subsection{Metals}
We have derived asymptotical analytical formulas Eqs. (\ref{sfinal}),(\ref{pfinal}) for the light absorptance in an opaque medium with a weakly rough surface at oblique incidence. They are correct provided that $\beta\ll 1, |\varepsilon|\beta^2\ll 1$. Below we present some physical results, which follow from these formulas. 
Let us consider, for example, the light absorptance of metals in the microwave region. In this case, $\varepsilon=4\pi i\sigma/\omega$, where $\sigma$ is almost equal to the dc conductivity and is a real number,$\varepsilon$ is a purely imaginary number.Taking into account that $|\varepsilon|\gg 1$, one finds from Eqs.(\ref{sfinal}),(\ref{pfinal})
\begin{equation}
A_{s,p}\approx A_{s,p0}\left(1+\frac{2\delta^2}{d^2}\right)
\label{micro}
\end{equation}
where $A_{s,p0}$ are absorptances of smooth surface for s- and p-polarizations, respectively, and $d=c/\sqrt{\pi\sigma\omega}$ is the skin depth \color{black} in metal at microwaves. This result qualitatively coincides with the one \cite{Tsang06} obtained previously.
 Fig.2 illustrates the normalized absorptance over a smooth silver surface in the microwave region. Here we used the expression for the silver DC conductivity from \cite{SilverDC}.
\color{black}

\begin{figure}
    \centering
    \begin{center}
    \vspace{2mm}
    \includegraphics[width=0.9\columnwidth]{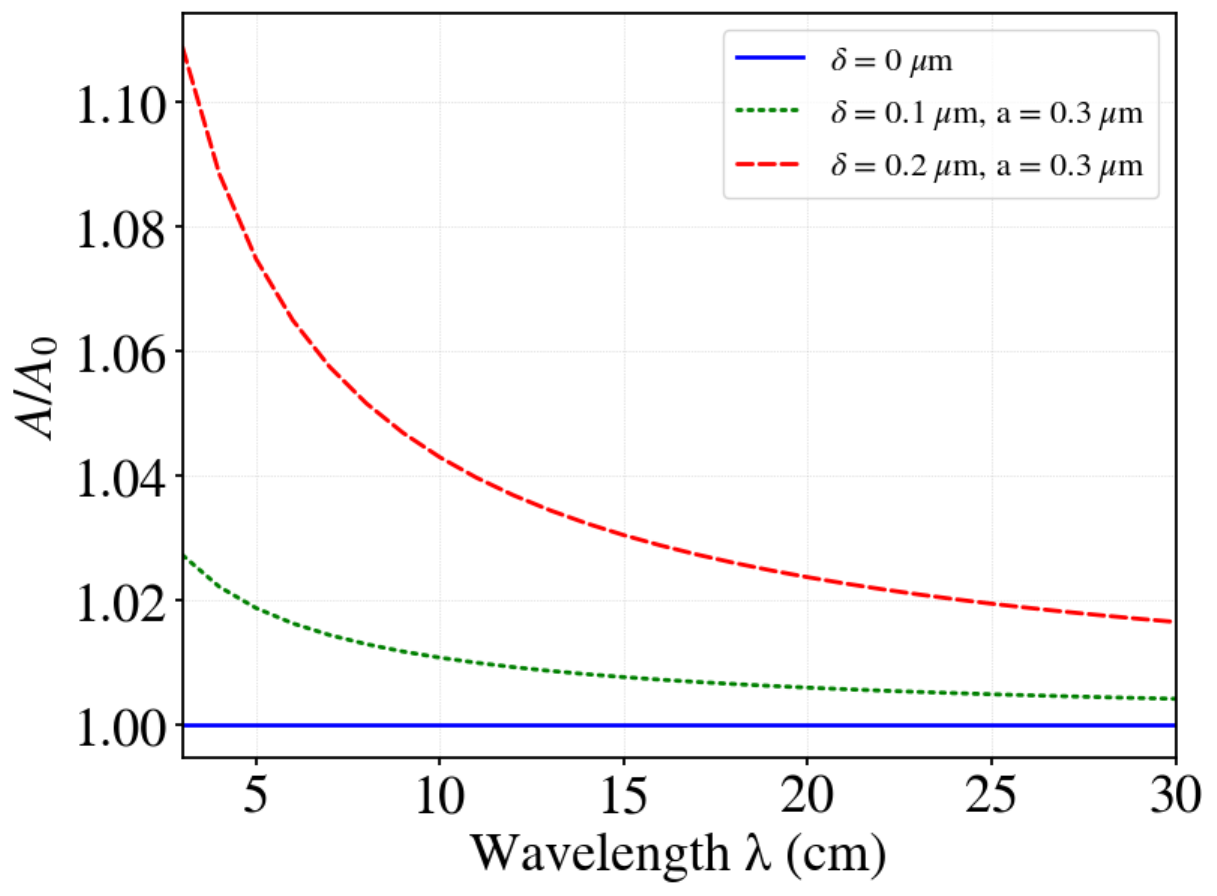}
    \caption{Light absorptance dependence on the wavelength in case of an oblique incidence of $50\degree$ on the silver surface.}     
    \end{center}
    \label{fig1.5}
\end{figure}

Differently, in the visible \color{black} and infrared the real part of $\varepsilon$ is a large negative number and it follows from Eqs. (\ref{sfinal}),(\ref{pfinal}) that all the roughness-induced corrections cancel each other provided that $|\varepsilon|\gg 1,|\sqrt{\varepsilon}|cos\theta\gg 1$, $A_{s,p}\approx A_{s,p0}$. Therefore, in the IR region the absorptance is insensitive to the roughness, as opposed to the MW region (see Figures 2,3 and 4 together with Eq.(\ref{micro}))
\begin{figure}[tb]
\centering
\begin{center}
\vspace{2mm}
\includegraphics[width=0.9\columnwidth]{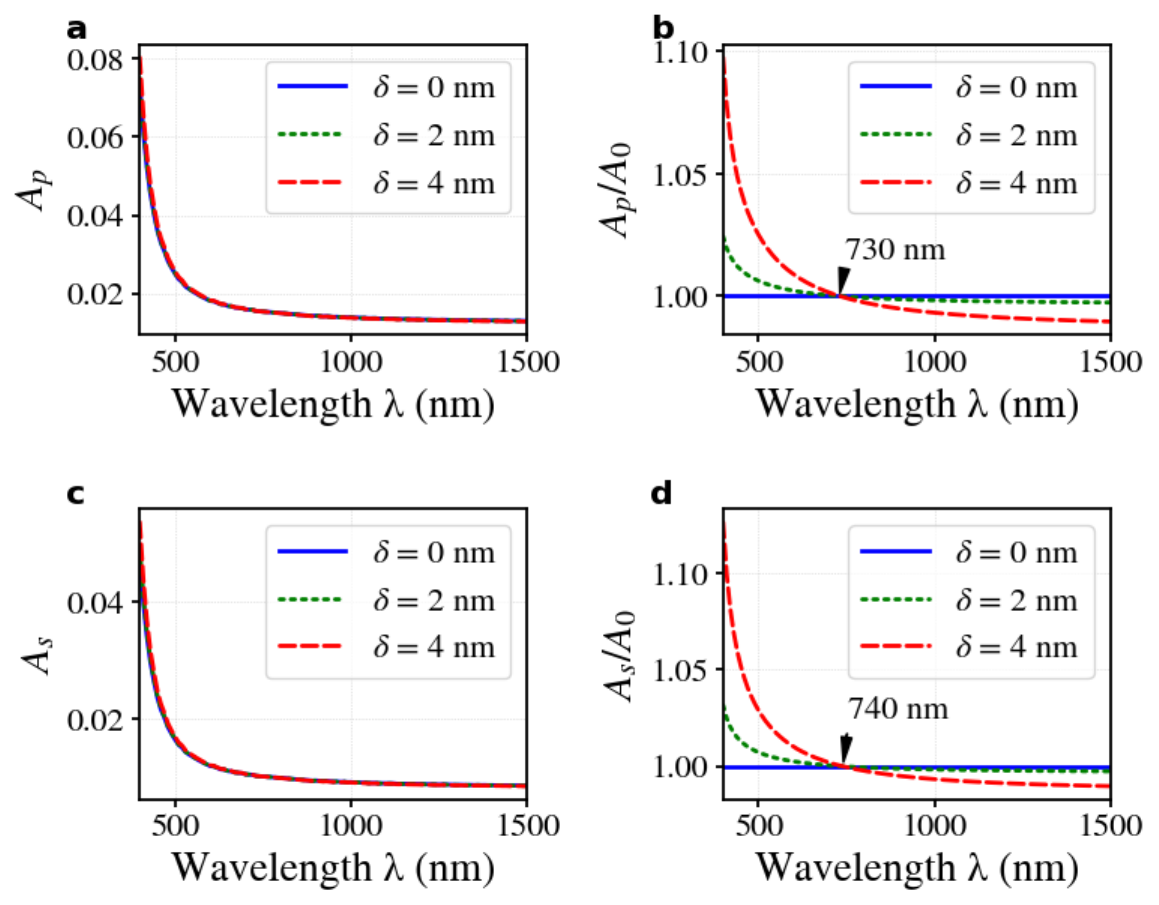}
\caption{Light absorptance dependence on its wavelength considering an oblique incidence $35\degree$ on silver surface. (a, c) Absolute absorptance and (b, d) normalized absorptance over smooth surface. In both cases, correlation length $a = 10 $ nm and different values of amplitude $\delta $ are considered, and $\delta = 0 $ nm corresponds to an ideally smooth surface. For an oblique incidence s-, (a, b), and p-, (c, d), polarization states are considered. The arrow in the inset indicates the wavelength at which the absorptance is equal for both the smooth and rough surfaces.}
\end{center}
\vspace{-6mm}
\label{fig2}
\end{figure}
 \begin{figure}[tb]
\centering
\begin{center}
\vspace{2mm}
\includegraphics[width=0.9\columnwidth]{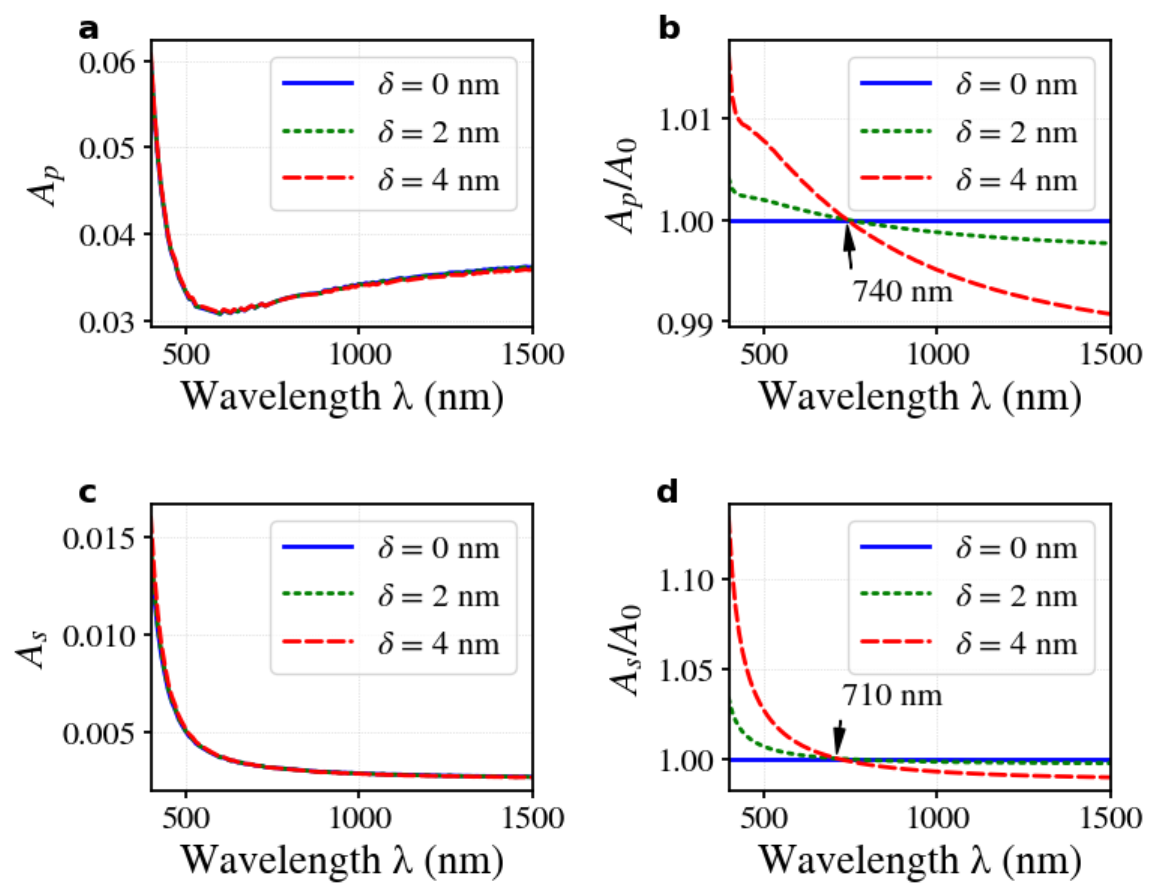}
\caption{The same plot as in Fig.3, but for an oblique incidence of  $75\degree$ \color{black}}
\end{center}
\vspace{-6mm}
\label{fig3}
\end{figure}

Below we present the results of numerical calculations of absorptance for weakly rough opaque silver films. The roughness parameters were chosen in the range $\delta\ll d$ and $a\lesssim d$, while the experimental dielectric function of silver was used in all calculations \cite{Jiang16}. The wavelength interval is chosen from $\lambda=400$ to 1500 nm in order to avoid the influence of surface plasmon ($\lambda\approx350$ nm in silver) and of interband transitions, both of which lead to enhanced absorption not directly related to roughness. 
All numerical calculations were carried out using the full expression for absorptance $A$, see the Appendix, while the small-scale roughness asymptotic expression Eqs. (\ref{sfinal}),(\ref{pfinal}) is used to discuss qualitative features of obtained results.

In Fig.3,4 \color{black} the relative absorptances for s- and p-polarized waves at incident angles $35\degree,75\degree$ in the optical region are shown. As was mentioned above, in this case, the roughness effect for the wavelengths $\lambda> 600$ nm is small and absorptances are almost equal to their values for the flat surface. The roughness effect is better seen in relative absorptance plots.
Due to the interference term absorptance is smaller than that for flat surfaces. Weak wavelength dependences of absorptances at $\lambda>600$ nm are associated with weak wavelength dependences of skin depth $d$ in this region \cite{yang15}. The absorptance value, as was mentioned above, is almost equal to that of a flat surface.
The increase of absorptance at $\lambda<400$ nm is caused by approaching the surface plasmon resonance \cite{raether88}(for silver $\sim 350$ nm) and interband transition. Plasmonic contribution to the absorptance is increasing with the roughness scale.

\subsection{Silicon in the visible and infrared regions}

In this section, we apply our formulae to the absorptance of rough silicon film in the optical region. We use the data for $n,\kappa$ from \cite{Green08}
\begin{figure}[tb]
\centering
\begin{center}
\vspace{2mm}
\includegraphics[width=0.9\columnwidth]{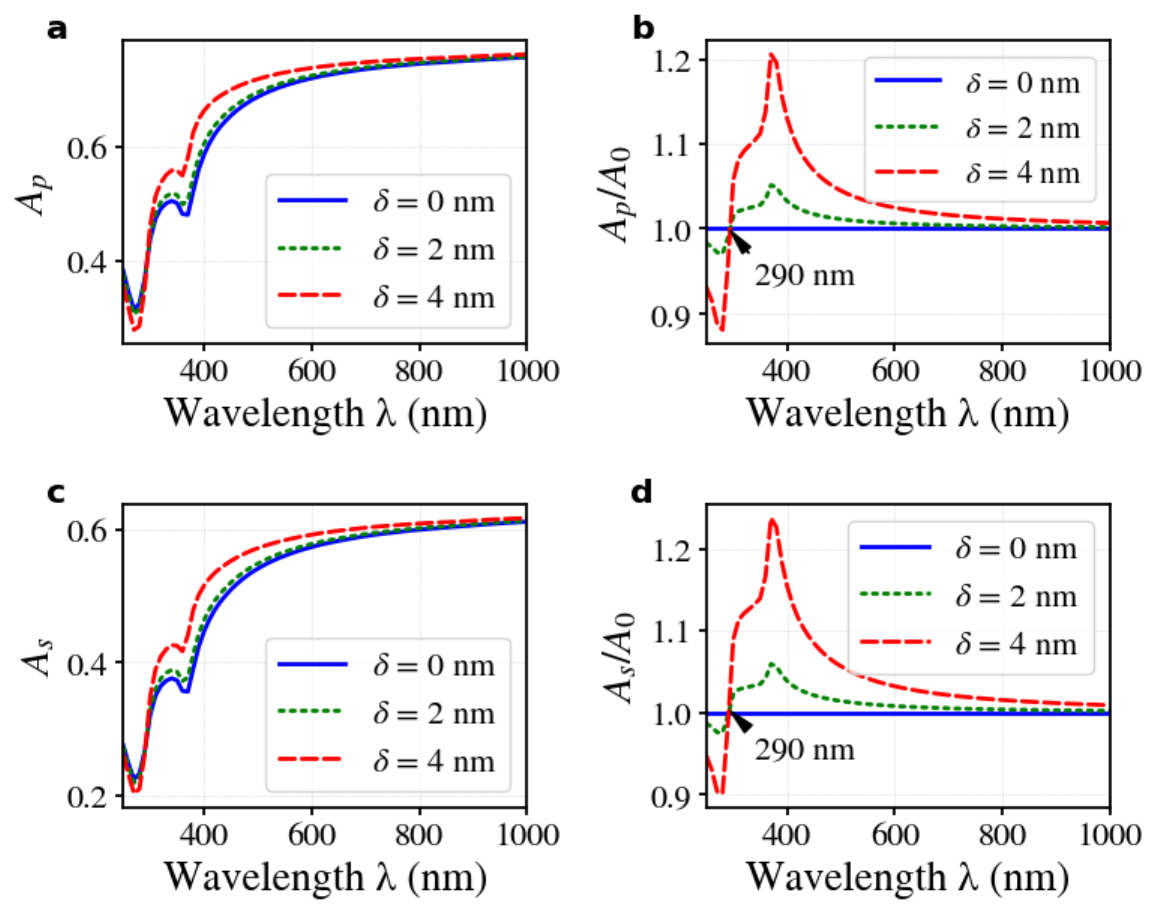}
\caption{Light absorptance dependence on its wavelength considering an oblique incidence $35\degree$ on the silicon surface. (a, c) Absolute absorptance and (b, d) normalized absorptance over smooth surface. In both cases, correlation length $a = 10 nm$ and different values of amplitude $\delta$ are considered, and $\delta = 0 nm$ corresponds to an ideally smooth surface. For an oblique incidence s-, (a, b), and p-, (c, d), polarization states are considered.}
\end{center}
\vspace{-6mm}
\label{fig4}
\end{figure}

As is seen from Fig.5 \color{black} roughness effect is significant for the wavelengths $\lambda<500$ nm. This is associated with approaching the direct band gap of silicon $\approx 360$ nm. At small incident angles absorptances for s- and p-polarizations are almost the same and roughness increases the absorptances at wavelengths $\lambda>290$ nm and decreases at wavelengths $\lambda<290$ nm. There is a wavelength of $290$ nm at which absorptances of rough and flat surfaces become equal. In an opaque medium, the reflectance is determined as $R=1-A$. At the above-mentioned wavelength, the reflectance from a rough surface is equal to the one from a flat surface. Therefore there will be no haze in the reflected light. At a small incident angle zero haze wavelengths coincide for s- and p-polarizations, see Fig.5 \color{black}. At an incident angle of $75\degree$, the zero haze wavelength for the s-polarized wave remains the same($290$ nm) while for the p-polarized wave, it moves to the larger wavelengths, see Fig.6 \color{black}.
\begin{figure}[tb]
\centering
\begin{center}
\vspace{2mm}
\includegraphics[width=0.9\columnwidth]{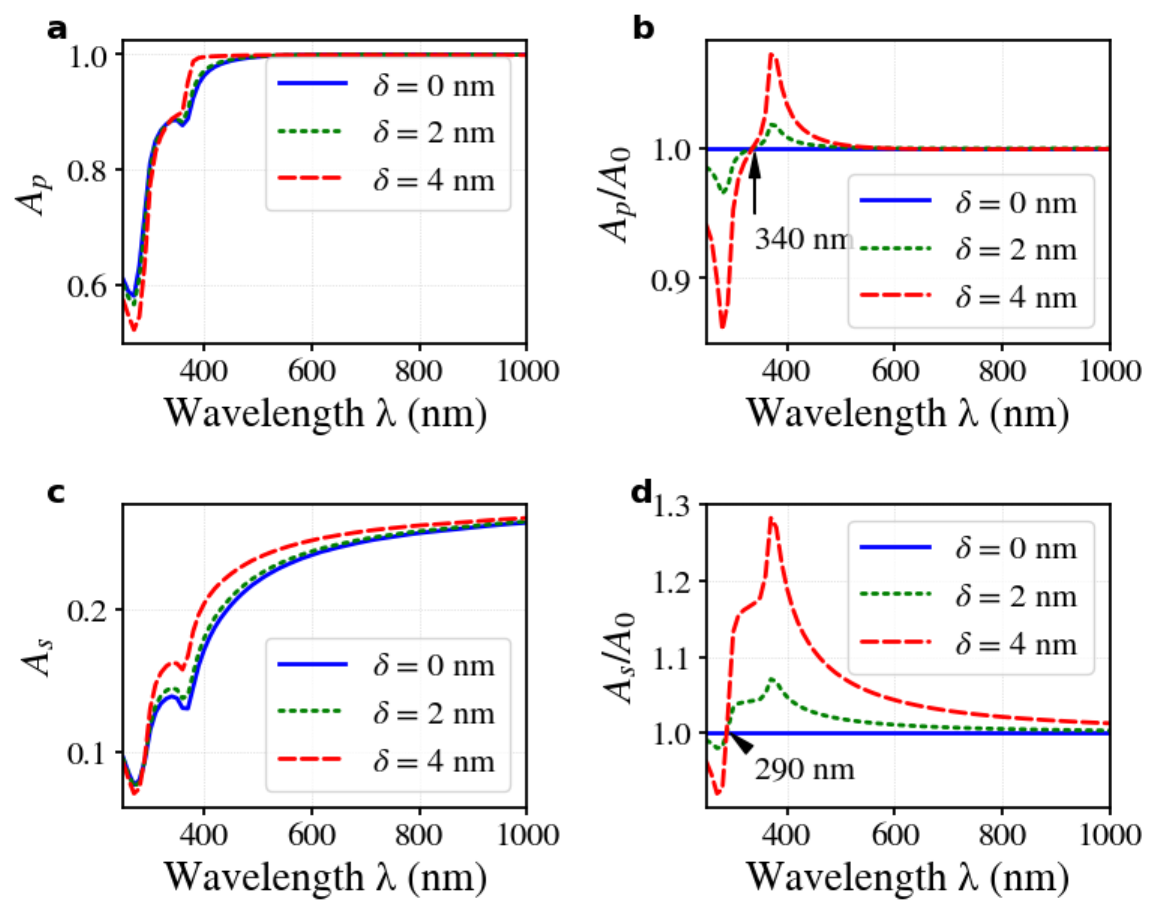}
\caption{The same plot as in Fig.5, but for an oblique incidence of  $75\degree$ \color{black}}
\end{center}
\vspace{25mm}
\label{fig5}
\end{figure}
 \begin{figure}[tb]
\centering
\begin{center}
\vspace{2mm}
\includegraphics[width=0.9\columnwidth]{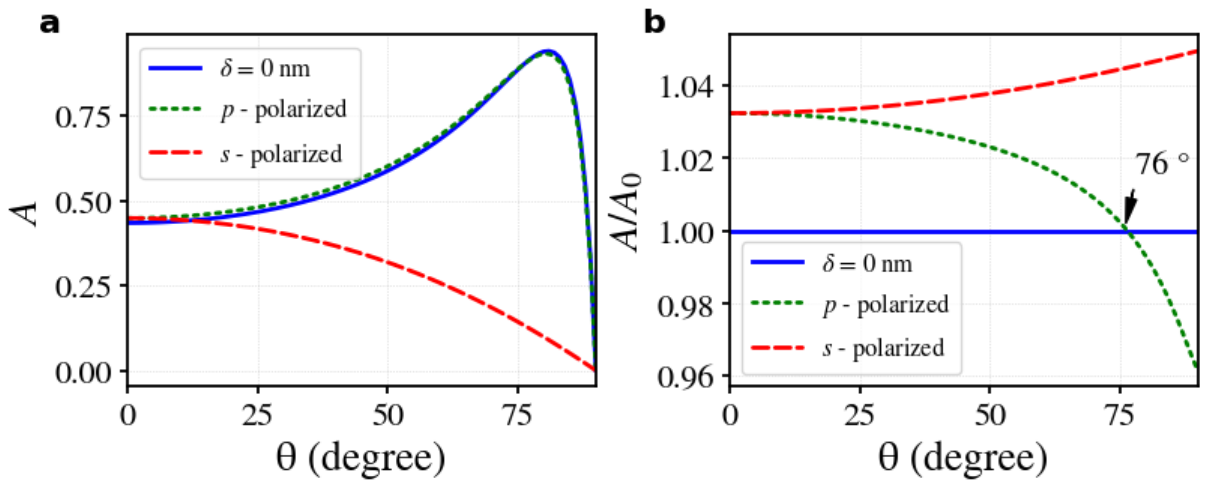}
\caption{Light absorptance dependence on the angle of incidence on the silicon surface. (a) Absolute absorptance and (b) normalized absorptance over a smooth surface. In both cases both s and p polarization states are considered, correlation length is $a = 10$ nm and the amplitude is $\delta =2$ nm, while $\delta = 0$ nm corresponds to an ideally smooth surface. The light wavelength is 350 nm. The arrow in the inset indicates the incident angle at which the absorptance is equal for both the smooth and rough surfaces}
\end{center}
\vspace{25mm}
\label{fig6}
\end{figure}
Figure 7 \color{black} shows the existence of an incident angle for which the absorptances and correspondingly the reflectances $R=1-A$ (there is no transmitted wave) for rough and flat surfaces are equal to each other. This means that, in terms of an integral reflectance, there will be no haze when a p-polarized wave is incident onto the rough surface at this angle. Consider now the reflectance as a sum of two parts: specular and non-specular(diffusive): $R(\theta)=R_s(\theta)+R_d(\theta)$. At zero haze incidence angle, the absorptance of a rough system equals that of a flat one: $R_s(\theta_{zh})+R_d(\theta_{zh})=R_f(\theta_{zh})$. As is shown in \cite{Navarette09}, $R_s(\theta_{zh})=R_f(\theta_{zh})(1-b\delta^2/\lambda^2)$ provided that $|\varepsilon|\gg 1$ and $b$ is a number of order unity. Substituting this equation into the above one obtains: $R_d(\theta_{zh})\sim R_f\delta^2/\lambda^2 \to 0$ for nanoscale roughness in visible or larger wavelengths region. On the contrary for other incident angles $R_d(\theta)\sim \delta^2/ad$.\color{black} 

Note that this zero haze angle differs from Brewster's angle at which absorptance is maximal. As seen from Fig.7 \color{black} for $\lambda=350nm$ zero haze angle equals $\theta_{zh}=76\degree$ while Brewster's angle for $\it Si$ at this wavelength equals to 
$80.6\degree$. Zero haze angle $\theta_{zh}$ depends on $a$ and not on $\delta$. Equalizing roughness-induced correction in Eq.(\ref{Apfinal}) to zero one can express the roughness correlation length through the $\theta_{zh}$
\begin{equation}
a=\frac{\sqrt{2\pi}}{4}\frac{Re\left[1-\frac{\varepsilon\cos\theta_{zh}}{i\kappa_{\theta}(\theta_{zh})(1+\sqrt{\varepsilon}\cos\theta_{zh})}\right]}{1-Re\left[\frac{\varepsilon}{\kappa^2_{\theta}(\theta_{zh})}\frac{1-\sqrt{\varepsilon}\cos\theta_{zh}}{1+\sqrt{\varepsilon}\cos\theta_{zh}}\right]}\times d(\theta_{zh})
\label{corlengthhaze}
\end{equation}
Here we use the asymptotic expression for $A_p$, Eq.(\ref{pfinal}).

It follows from Fig.5,6 \color{black} that for short wavelengths $\lambda<300nm$ the zero haze angle exists also for the s-polarized wave contrary to Brewster's angle which exists only for the p-polarized wave.

 As to the Brewster angle, it shifts because of surface roughness \cite{saillard90}-\cite{mar93}.
\section{4. Summary}
\label{sec-conc}

In summary, we developed a perturbative approach for absorption of oblique incident light in weakly rough opaque systems characterized by a Gaussian surface profile with RMS amplitude $\delta$ and correlation length $a$ that are smaller than penetration depth $d$ in the medium. General formulae are derived that apply to all materials and wavelength regions. We have shown, that in such systems the accurate choice of boundary conditions leads to a negative contribution of interference term that dominates in certain cases. Polarization and angular dependences of absorptance are revealed. We demonstrate the existence of zero haze incidence angle and wavelength at which absorptance and reflectance $R=1-A$ equal to those of a flat surface. This phenomenon allows us to independently determine the Gaussian roughness correlation length from optical measurements.

 For the calculated average absorptance to have any relation to the experiment it should be self-averaged, namely, the sample-to-sample fluctuations should be small. We can consider our system to consist of a large number of macroscopic parts. If we assume that the integral absorptance is an ergodic quantity, averaging over the ensemble can
be substituted by averaging over different parts. Besides, the fluctuations in the average value because of the central limit theorem
will be inversely proportional to the square root of the number of constituent
parts and thus, can be made negligibly small. This puts our theoretically derived averaged quantity to that of the experimental one. 


\section{Acknowledgments} This work was supported by Armenia Science Committee Grant  No. 21AG-1C062.

\section{Appendix}

\begin{appendix}

\subsection{A: General expression for the absorption ratio in case of S-polarized wave}
\renewcommand{\theequation}{A.\arabic{equation}}
\setcounter{equation}{0}
\label{sec-appen}
\begin{eqnarray}
    \frac{A_s}{A_{s0}}=1+ \frac{2\delta^2}{d^2}-2k_0^2 \delta^2 Re\left[(\varepsilon-1)\times \right.\nonumber\\ \left. \times \frac{\cos\theta-\sqrt{\varepsilon-\sin^2\theta}}{\cos\theta+\sqrt{\varepsilon-\sin^2\theta}}\right]+ \nonumber\\ +\frac{4\delta^2}{ad} Re[(\varepsilon-1)I^{2D}(k_0a)]
    -\frac{2\delta^2}{ad}\times \nonumber\\ \times Re\left[\frac{(\varepsilon-1)^2 I^{2D}(k_0a)}{i\kappa_{\theta}(\cos\theta+\sqrt{\varepsilon-\sin^2\theta})}\right]+\nonumber\\ +\frac{2\delta^2}{ad}|\varepsilon-1|^2 I^{2D}_s(k_0a)
\end{eqnarray}
with
\begin{flalign}
I^{2D}(\beta)=i \times exp(-\beta^2 \sin^2\theta / 2) \int_{0}^{\infty} dr r \times exp(-r^2/2) \times \nonumber\\ \left[ \frac{\sqrt{(\varepsilon \beta^2-r^2)(\beta^2-r^2)}}{\sqrt{\varepsilon \beta^2-r^2}+\varepsilon \sqrt{\beta^2-r^2}} 
f_{+}(r \beta \sin\theta)+ \right.\nonumber\\ \left.+\frac{\beta^2}{\sqrt{\varepsilon \beta^2-r^2}+\sqrt{\beta^2-r^2}} 
f_{-}(r \beta \sin\theta) \right] \nonumber\\
\label{I2DApp}
\end{flalign}

\begin{flalign}
I^{2D}_s(\beta)=exp(-\beta^2 \sin^2\theta / 2) \int_{0}^{\infty} dr r \frac{exp(-r^2/2)}{2|Im\sqrt{\varepsilon \beta^2-r^2}|} \times  \nonumber\\
\left[ \frac{|\beta^2-r^2|(r^2+|\varepsilon \beta^2-r^2|)}{|\sqrt{\varepsilon \beta^2-r^2}+\varepsilon \sqrt{\beta^2-r^2}|^2} 
f_{+}(r \beta \sin\theta)+\right.\nonumber\\ \left.+ \frac{\beta^4}{|\sqrt{\varepsilon \beta^2-r^2}+\sqrt{\beta^2-r^2}|^2} 
f_{-}(r \beta \sin\theta) \right]\nonumber\\
\label{I2DsApp}
\end{flalign}
where
\begin{equation}
    f_{\pm}(x)=\frac{I_0(x) \mp I_2(x)}{2}
    \label{f}
\end{equation}
\subsection{B: Derivation of the absorption ratio for the P-polarized case}
\renewcommand{\theequation}{B.\arabic{equation}}
\setcounter{equation}{0}

For brevity, we omit the index in $r_p$ and $t_p$. The only non-zero components of an electric field are
\begin{align}
\tilde{E}_{0x}({\bf r})=\left\{\begin{array}{c}\cos\theta \times (e^{ik_xx-ik_{z}z}-re^{ik_xx+ik_{z}z}),\\ \mbox{~for~} z>h(x,y),
\\ \\
t\cos\theta_t \times e^{ik_xx+ik_{-}z}, \\ \mbox{\qquad\qquad for~} z<h(x,y),
\end{array}\right.
\label{backx}
\end{align}
and
\begin{align}
\tilde{E}_{0z}({\bf r})=\left\{\begin{array}{c}\sin\theta \times(e^{ik_xx-ik_{z}z}+re^{ik_xx+ik_{z}z}),\\ \mbox{~for~} z>h(x,y),
\\ \\
t\sin\theta_t \times e^{ik_xx+ik_{-}z}, \\ \mbox{\qquad\qquad for~} z<h(x,y),
\end{array}\right.
\label{backz}
\end{align}
,where  $\sin\theta_t=\sin\theta/\sqrt{\varepsilon}, \cos\theta_t=\sqrt{1-\sin^2\theta/\varepsilon}$, $\theta_t$ is the transmission angle and is generally complex and $k_{-}=-k_{z-}$.

\subsubsection{Reference field contribution}

Putting Eqs. (\ref{backx},\ref{backz}) into Eq. (\ref{bulbac}) we obtain 
\begin{equation}
A_{pr}=\frac{\varepsilon''k_{0}|T|^2}{S_{0}\cos\theta}\left\langle\int dxdy \int_{-\infty}^{h(x,y)}e^{-2\kappa_{\theta}k_0z}\right\rangle.
\label{bulbacp}
\end{equation}
,where $T=t\times \tau, \tau=\sqrt{|1-\sin^2\theta/\varepsilon|+\sin^2\theta/|\varepsilon|}$. 
With $A_{p0}=\varepsilon''|T|^2/2\kappa_{\theta}\cos\theta$ we write
\begin{equation}
A_{pr}=A_{p0}\left[1+\frac{2\delta^2}{d^2}\right]
\label{bulbacp1}
\end{equation}

\subsubsection{Interference term contribution}

As previously for the S-polarization case, by putting Eqs.(\ref{scatf},\ref{backx},\ref{backz}) into Eq.(\ref{intterm}) we get
\begin{eqnarray}
A_i^1=-\frac{k_0^3\varepsilon'' (\varepsilon-1)}{2\pi S_0 \cos\theta}|t|^2
Re\left\langle\int d{\vec \rho}d{\vec \rho'}\int_{-\infty}^hdz\times \right.\nonumber\\
\left. \times \int\frac{d{\vec q}}{(2\pi)^2}e^{i({\vec q}-k_x\vec e_x)({\vec \rho}-{\vec \rho'})}\times \right.\nonumber\\
\left.\times \left( \sum_{\alpha,\alpha^{\prime}\in \{x,z\}}[\textbf{d}(\vec q,z,h(\vec{\rho^{\prime}}))]_{\alpha,\alpha^{\prime}}n^{*}_{\alpha} n_{\alpha^{\prime}} \right)\times \right.\nonumber\\
\left. \times e^{ik_{z_{-}}h(\vec{\rho^{\prime}})-ik_{z_{-}}^*z}h(\vec{\rho^{\prime}})\right\rangle
\label{intterm2p}
\end{eqnarray}
,where $\alpha, \alpha^{\prime}$ in sum take values of either $x$ or $z$ independently from each other and $n_x=\sin \theta/(\sqrt{\varepsilon}\tau)$, $n_z=\tau^{-1}\sqrt{1-\sin^2 \theta/ \varepsilon}$.

Averaging by using $<h(\Vec{\rho})h(\Vec{\rho^{\prime}})>=\delta^2exp(-(\Vec{\rho}-\Vec{\rho^{\prime}})^2/2a^2)$, then integrating alternatively we get
\begin{eqnarray}
    A_i^1=\frac{4A_0 \delta^2}{ad}Re\left[(\varepsilon-1)\sum_{\alpha,\alpha^{\prime}\in \{x,z\}}I^{(2D)}_{\alpha,\alpha^{\prime}}(k_0 a) n^{*}_{\alpha} n_{\alpha^{\prime}}\right] \nonumber \\ +2A_0\delta^2 Re\left[(\varepsilon-1)\sum_{\alpha,\alpha^{\prime}\in \{x,z\}}H_{\alpha,\alpha^{\prime}} n^{*}_{\alpha} n_{\alpha^{\prime}}\right]\nonumber\\
    \label{inttermp1}
\end{eqnarray}

, where 
\begin{eqnarray}
    I^{(2D)}_{xx}(\beta)= ie^{-\beta^2 \sin^2 \theta/2} \int_0^{\infty}dr r e^{-r^2/2} \nonumber \\ \times \left[ \frac{\sqrt{\beta^2-r^2}\sqrt{\varepsilon \beta^2-r^2}}{\sqrt{\varepsilon \beta^2-r^2}+\varepsilon \sqrt{\beta^2-r^2}} f_{-}(r\beta\sin \theta)\right.\nonumber\\
\left. + \frac{\beta^2}{\sqrt{\varepsilon \beta^2-r^2}+\sqrt{\beta^2-r^2}} f_{+}(r\beta\sin \theta) \right] \nonumber \\
    I^{(2D)}_{xz}(\beta)=ie^{-\beta^2 \sin^2 \theta/2} \int_0^{\infty}dr r e^{-r^2/2}\nonumber\\
    \left[\frac{r\sqrt{\varepsilon\beta^2-r^2}}{\varepsilon(\sqrt{\varepsilon \beta^2-r^2}+\varepsilon \sqrt{\beta^2-r^2})}I_1(r\beta\sin \theta)\right] \nonumber \\
    I^{(2D)}_{zx}(\beta)=ie^{-\beta^2 \sin^2 \theta/2} \int_0^{\infty}dr r e^{-r^2/2}\nonumber\\
    \left[\frac{r\sqrt{\beta^2-r^2}}{\sqrt{\varepsilon \beta^2-r^2}+\varepsilon \sqrt{\beta^2-r^2}}I_1(r\beta\sin \theta)\right] \nonumber \\
    I^{(2D)}_{zz}(\beta)=ie^{-\beta^2 \sin^2 \theta/2} \int_0^{\infty}dr r e^{-r^2/2}\nonumber\\
    \left[\frac{r^2}{\varepsilon(\sqrt{\varepsilon \beta^2-r^2}+\varepsilon \sqrt{\beta^2-r^2})}I_0(r\beta\sin \theta)\right]
    \label{Ixxxzzxzz}
\end{eqnarray}

\begin{eqnarray}
    H_{xx}=-k^2_{-}\frac{k_{-}/\varepsilon+k_0\cos \theta}{k_{-}-\varepsilon k_0\cos \theta} \nonumber \\
    H_{xz}=-k_{-}k_0\sin \theta \frac{k_{-}/\varepsilon+k_0\cos \theta}{k_{-}-\varepsilon k_0\cos \theta} \nonumber \\
    H_{zx}=k_{-}k_0\sin \theta \frac{k_{-}/\varepsilon+k_0\cos \theta}{k_{-}-\varepsilon k_0\cos \theta} \nonumber \\
    H_{zz}=k^2_{0}\sin^2 \theta \frac{k_{-}/\varepsilon+k_0\cos \theta}{k_{-}-\varepsilon k_0\cos \theta}
    \label{Hxxxzzxzz}
\end{eqnarray}

For the second interference contribution $A_{i2}$ substitute Eqs.(\ref{seconord},\ref{backx},\ref{backz}) into Eq.(\ref{intterm}). The successive evaluation leads to

\begin{eqnarray}
    A_i^2=\frac{2 A_0 \delta^2}{a} Re\left[(\varepsilon-1)^2 \times
\right.\nonumber\\
\left. \times \sum_{\alpha,\alpha^{\prime}\in \{x,z\}}\left(\textbf{G} \times \textbf{I}^{(2D)}(k_0 a)\right)_{\alpha,\alpha^{\prime}} n^{*}_{\alpha} n_{\alpha^{\prime}} \right]
    \label{inttermp2}
\end{eqnarray}
,where $\left(\textbf{G} \times \textbf{I}^{(2D)}\right)_{\alpha,\alpha^{\prime}}=\sum_{\gamma \in \{x,z\}}G_{\alpha,\gamma}I^{(2D)}_{\gamma,\alpha^{\prime}}$ and
\begin{eqnarray}
    G_{xx}=ik_0\frac{\cos \theta\sqrt{\varepsilon-\sin^2 \theta}}{\sqrt{\varepsilon-\sin^2 \theta}+ \varepsilon \cos \theta}\nonumber\\
    G_{xz}=ik_0\frac{\sin \theta\sqrt{\varepsilon-\sin^2 \theta}}{\varepsilon(\sqrt{\varepsilon-\sin^2 \theta}+ \varepsilon \cos \theta)}\nonumber\\
    G_{zx}=ik_0\frac{\sin \theta \cos \theta }{\sqrt{\varepsilon-\sin^2 \theta}+ \varepsilon \cos \theta}\nonumber\\
    G_{zz}=ik_0\frac{\sin^2 \theta}{\varepsilon (\sqrt{\varepsilon-\sin^2 \theta}+ \varepsilon \cos \theta)}
    \label{Gxxxzzxzz}
\end{eqnarray}

\subsubsection{Scattering term contribution}

Substituting Eqs.(\ref{scatf},\ref{backx},\ref{backz}) into Eq.(\ref{bulkscat}) and performing the successive evaluations we get

\begin{eqnarray}
    A_{sc}=\frac{2A_0 \delta^2}{ad}|\varepsilon-1|^2 \sum_{\alpha,\alpha^{\prime}\in \{x,z\}}(I_s)^{(2D)}_{\alpha,\alpha^{\prime}}(k_0 a) n^{*}_{\alpha} n_{\alpha^{\prime}}\nonumber\\
    \label{scattermp}
\end{eqnarray}
, where 
\begin{eqnarray}
    (I_s)^{(2D)}_{xx}(\beta)=e^{-\beta^2 \sin^2 \theta/2}\int_0^{\infty}dr r \frac{e^{-r^2/2}}{2Im[\sqrt{\varepsilon \beta^2-r^2}]}\nonumber\\ \times \left[ \frac{|\beta^2-r^2|(r^2+|\varepsilon\beta^2-r^2|)}{|\sqrt{\varepsilon\beta^2-r^2}+\varepsilon\sqrt{\beta^2-r^2}|^2} f_{-}(r\beta \sin \theta)+\right.\nonumber\\
\left.+\frac{\beta^4}{|\sqrt{\varepsilon\beta^2-r^2}+\sqrt{\beta^2-r^2}|^2} f_{+}(r\beta \sin \theta)\right]\nonumber\\ \nonumber\\
    (I_s)^{(2D)}_{xz}(\beta)=e^{-\beta^2 \sin^2 \theta/2}\int_0^{\infty}dr r^2 \frac{e^{-r^2/2}}{2Im[\sqrt{\varepsilon \beta^2-r^2}]}\nonumber\\ \times \left[ \frac{(r^2+|\varepsilon\beta^2-r^2|)(\sqrt{\beta^2-r^2})^{*}}{\varepsilon|\sqrt{\varepsilon\beta^2-r^2}+\sqrt{\beta^2-r^2}|^2} I_1(r\beta\sin\theta)\right]\nonumber\\ \nonumber\\
    (I_s)^{(2D)}_{zx}(\beta)=((I_s)^{(2D)}_{xz}(\beta))^{*}\nonumber\\ \nonumber\\
    (I_s)^{(2D)}_{zz}(\beta)=e^{-\beta^2 \sin^2 \theta/2}\int_0^{\infty}dr r^3 \frac{e^{-r^2/2}}{2Im[\sqrt{\varepsilon \beta^2-r^2}]}\nonumber\\ \times \left[\frac{r^2+|\varepsilon\beta^2-r^2|}{|\varepsilon|^2|\sqrt{\varepsilon\beta^2-r^2}+\varepsilon\sqrt{\beta^2-r^2}|^2} I_0(r\beta\sin\theta)\right]\nonumber\\
    \label{Isxxxzzxzz}
\end{eqnarray}

Finally, the full absorptance for p-wave:
\begin{eqnarray}
    \frac{A_p}{A_{p0}}=1+\frac{2\delta^2}{d^2}+\frac{4 \delta^2}{ad}Re\left[(\varepsilon-1)\right.\nonumber\\
\left.\sum_{\alpha,\alpha^{\prime}\in \{x,z\}}I^{(2D)}_{\alpha,\alpha^{\prime}}(k_0 a) n^{*}_{\alpha} n_{\alpha^{\prime}}\right]  +2\delta^2  Re[(\varepsilon-1)\nonumber \\ \times \sum_{\alpha,\alpha^{\prime}\in \{x,z\}}H_{\alpha,\alpha^{\prime}} n^{*}_{\alpha} n_{\alpha^{\prime}}] + \frac{2 \delta^2}{a} Re\left[(\varepsilon-1)^2 
\right.\nonumber\\
\left. \sum_{\alpha,\alpha^{\prime},\gamma\in \{x,z\}}\left(G_{\alpha,\gamma}I^{(2D)}_{\gamma,\alpha^{\prime}}(k_0 a)\right)_{\alpha,\alpha^{\prime}} n^{*}_{\alpha} n_{\alpha^{\prime}} \right]\nonumber\\ +\frac{2 \delta^2}{ad}|\varepsilon-1|^2   \sum_{\alpha,\alpha^{\prime}\in \{x,z\}}(I_s)^{(2D)}_{\alpha,\alpha^{\prime}}(k_0 a) n^{*}_{\alpha} n_{\alpha^{\prime}}\nonumber\\
    \label{Apfinal}
\end{eqnarray}

\end{appendix}

\end{document}